\documentclass[conference]{IEEEtran}
\IEEEoverridecommandlockouts
\usepackage{changes}
\definechangesauthor[name={vahid}, color=red]{1}

\usepackage{cite}
\usepackage{amsmath,amssymb,amsfonts}
\usepackage{mathrsfs}
\usepackage{algorithmic}
\usepackage{graphicx}
\usepackage{textcomp}
\usepackage{xcolor}
\usepackage{url}
\usepackage{float}
\usepackage{amsmath}

\DeclareMathOperator*{\argmin}{arg\,min}
\usepackage{hyperref}

\def\BibTeX{{\rm B\kern-.05em{\sc i\kern-.025em b}\kern-.08em
    T\kern-.1667em\lower.7ex\hbox{E}\kern-.125emX}}
\begin{document}

	\title{\textbf{Deep Learning-Based Channel Estimation}}
    
     \author{ \IEEEauthorblockN{Mehran Soltani,  Vahid Pourahmadi, Ali Mirzaei, Hamid Sheikhzadeh}}
     		
    	
    \maketitle

\begin{abstract}
	In this paper, we present a deep learning (DL) algorithm for channel estimation in communication systems. We consider the time-frequency response of a fast fading communication channel as a two-dimensional image. The aim is to find the unknown values of the channel response  using some known values at the pilot locations. To this end, a general pipeline using deep image processing techniques, image super-resolution (SR) and image restoration (IR) is proposed. This scheme considers the pilot values, altogether, as a low-resolution image and uses an SR network cascaded with a denoising IR network to estimate the channel. Moreover, an implementation of the proposed pipeline is presented. The estimation error shows that the presented algorithm is comparable to the minimum mean square error (MMSE) with full knowledge of the channel statistics and it is better than ALMMSE (an approximation to linear MMSE). The results confirm that this pipeline can be used efficiently in channel estimation.

\end{abstract}


\section{Introduction}
 Orthogonal frequency-division multiplexing (OFDM) is a modulation method that has been widely used in communication systems to address frequency-selective fading in wireless channels. In a communication channel, the received signal is usually distorted by channel characteristics. In order to recover the transmitted symbols, the channel effect must be estimated and compensated at the receiver. Generally, the receiver estimates the channel using some symbols named pilots which their positions and values in time-frequency are known to both transmitter and receiver. Depending on these pilot arrangements, three different structures can be considered: block-type, comb-type and lattice-type \cite{CombPilot}. In the block-type arrangement, pilots are transmitted periodically at the beginning of an OFDM block at all subcarriers while in comb-type, the pilots are present in few subcarriers of few OFDM symbols. In the lattice-type arrangement, pilots are inserted along both time and frequency axes with given periods in a diamond-shaped constellation. 

The conventional pilot-based estimation methods, i.e., Least Square (LS) and Minimum Mean Square Error (MMSE) utilize the pilot values in time-frequency grids to find the unknown values of the channel response. These algorithms have been optimized in various conditions \cite{Lichannel}. In contrast to the LS estimation which requires no information about the statistics of the channel, the MMSE estimation results in a better performance by utilizing the statistics of the channel and noise variance. To use the MMSE in practical scenarios, some approaches are presented which reduce the complexity of this scheme and use an estimation of the channel statistics instead of the exact information.
In \cite{ALMMSE_CO}, an approximated linear version of the MMSE (ALMMSE), is proposed in fast fading channels which its complexity is much less than the original MMSE due to reducing the size of the correlation and the filtering matrix.

Recently, Deep Learning (DL) has gained much attention in communication systems. In DL-based communication systems, some approaches have been proposed to enhance the performance of different conventional algorithms including modulation recognition \cite{DBLPOShea}, signal detection  \cite{8227772}, channel equalization \cite{943148}, channel state information (CSI) feedback \cite{DBLP1712-08919} and channel estimation \cite{ye2018power},\cite{Deepchannel}. In \cite{ye2018power}, the communication system is considered as a black-box and an end-to-end DL architecture is used for signal transmission/reception. Encoding, decoding, channel estimation and all other functionalities of a communication link are embedded in the DL-block, implicitly. More specifically, this method is not able to explicitly find the channel time-frequency response and so not effective for applications which need to have the complete channel response. In \cite{Deepchannel}, the channel matrix is considered as an image and then used a denoising network for channel estimation. This work focuses on the channel matrix along the transmitter/receiver antenna space (in multiple antenna scenario) and is not discussing the time-frequency response of the each Tx/Rx link.      


Motivated by this, in this paper, we present a DL-based framework for channel estimation in OFDM systems.  In this method, the time-frequency grid of the channel response is modeled as a 2D-image which is known only at the pilot positions. This channel grid with several pilots is considered as a low-resolution (LR) image and the estimated channel as a high-resolution (HR) one. A two-phase approach is presented to estimate the channel grid. First, an image super-resolution (SR) algorithm is used to enhance the resolution of the LR input. Afterwards, an image restoration (IR) method is utilized to remove the noise effects. For SR and IR networks, we have used two recently developed CNN-based (Convolutional Neural Network) algorithms, SRCNN \cite{SRCNN} and DnCNN \cite{ِDenoising}, respectively. The contributions of this paper are summarized as follows:
		\begin{enumerate}
			\item Model the channel time-frequency response as an image.
			\item Consider the channel response in the pilot positions as a LR image and the estimated channel response as the proposed HR image.
			\item Use DL-based image super-resolution and image denoising techniques to estimate the channel.
		\end{enumerate}  
The remainder of the paper is organized as follows. Section \ref{sec:ch} provides a brief survey of the channel estimation with conventional methods. Section \ref{sec:chnet} presents the structure of the proposed DL-base channel estimator. In section \ref{sec:sim}, simulation results are presented and finally section \ref{sec:conc} concludes the paper.

\section{Background} \label{sec:ch}

\subsection{Channel Estimation} 

     In an OFDM system, for the $k$th \textcolor{black}{time slot} and the $i$th subcarrier, the input-output relationship is represented as:
     
     \begin{equation} \label{eq:ch_mo}
     Y_{i,k}=H_{i,k}X_{i,k} +Z_{i,k}.    
     \end{equation}  

     Considering an OFDM \textcolor{black}{subframe} of size $N_S \times N_D$, \textcolor{black}{time slot} index $k$ is between $[0,N_D-1]$ and the range of the subcarrier index $i$ is $[0,N_S-1]$.
     In \eqref{eq:ch_mo}, $Y_{i,k}$, $X_{i,k}$, and $Z_{i,k}$ are the received signal, transmitted OFDM symbol and white Gaussian noise, respectively. $H_{i,k}$ is the $(i,k)$ element of $\mathbf{H}\in \mathbb{C}^{N_S \times N_D}$. $\mathbf{H}$ represents time-frequency response of the channel for all subcarriers  and \textcolor{black}{time slots}. 
     
     To estimate the channel, specifically in the channels with fading, the time domain response is \textcolor{black}{represented as} $\mathbf{H}=\{\mathbf{h}[1],\mathbf{h}[2],...,\mathbf{h}[N_D]\} $, where each $\mathbf{h}[k]$ is the channel frequency response at the $k$th \textcolor{black}{time slot}. 
     
      The LS method estimates the channel at the pilot positions. If we consider the LS estimated channel as a diagonal matrix $\mathbf{H}_p^{\mathrm{LS}} \in \mathbb{C}^{N_P \times N_P}$, $\mathbf{H}_p^{\mathrm{LS}}$ can be estimated by solving:     
     \begin{equation} \label{eq:LS}
     \mathbf{\hat{H}}_p^{\mathrm{LS}} = \argmin_{\mathbf{H}_p}  \|\mathbf{y}_p-\mathbf{H}_p\mathbf{x}_p\|_2^2,
     \end{equation}
     where $||.||_2$ is the $\ell2$ distance and $\mathbf{\hat{H}}_p^{\mathrm{LS}} \in \mathbb{C}^{N_P \times N_P}$ is the estimated diagonal matrix. $\mathbf{x}_{p}$ contains the known pilot values and $\mathbf{y}_{p}$  is the corresponding observations. The optimization of \eqref{eq:LS} results in $\mathbf{\hat{h}}_{p}^{\mathrm{LS}}=\mbox{diag}(\mathbf{\hat{H}}_p^{\mathrm{LS}})=\mathbf{y}_p/\mathbf{x}_p$. To find the channel value at the points other than the pilot positions, we have to apply a two dimensional interpolation method.
     
     
     A better choice than LS, is MMSE estimator which is obtained by multiplying the LS estimates at the pilot-symbol positions with a filtering matrix \textcolor{black}{$\textbf{A}_{\mathrm{MMSE}} \in \mathbb{C}^{N_L \times N_P}$ } \cite{Omar}:
     
     \begin{equation}
     	\hat{\textbf{h}}_{d}^{\mathrm{MMSE}}=\textbf{A}_{\mathrm{MMSE}} \hat{\textbf{h}}_{p}^{\mathrm{LS}},
     \end{equation}
     where \textcolor{black}{$\hat{\textbf{h}}_{d}^{\mathrm{MMSE}} \in \mathbb{C}^{N_L \times 1}$} ($N_L=N_S\times N_D$) is the \textcolor{black}{vectorized} MMSE estimation of the channel response \textcolor{black}{ $\mathbf{H}$} at subframe $d$. To find the filtering matrix, the mean square error(MSE),
     
     \begin{equation} \label{eq:MSE}
      \epsilon = \mathbb{E}\{{\|\mathbf{h}_d-\textbf{A}_{\mathrm{MMSE}}\hat{\textbf{h}}_{p}^{\mathrm{LS}}\|_2^2}\},
     \end{equation}
     has to be minimized. Minimizing \eqref{eq:MSE} leads to 
     \begin{equation}
     \textbf{A}_{\mathrm{MMSE}} = \mathbf{R}_{\mathbf{h}_{d}\mathbf{h}_{p}}(\mathbf{R}_{\mathbf{h}_{p}\mathbf{h}_{p}}+
     \sigma_{n}^2(\mathbf{x}\mathbf{x}^{\mathrm{H}})^{-1})^{-1}, 
     \end{equation}
	where the matrix $\mathbf{R}_{\mathbf{h}_{d}\mathbf{h}_{p}} = \mathbb{E}\{\mathbf{h}_{d}\mathbf{h}_{p}^{\mathrm{H}}\} 
	$
	denotes 
	the channel correlation matrix between desired subframe and pilot-symbols and the matrix $\mathbf{R}_{\mathbf{h}_{p}\mathbf{h}_{p}} =\mathbb{E}\{\mathbf{h}_{p}\mathbf{h}_{p}^{\mathrm{H}}\} 
	$
	is the channel correlation matrix at the pilot-symbols.
	It is obvious that the MMSE will be useful only if the correlation matrix of the channel, denoted as $\mathbf{R}$,  is completely known.  

	\subsection{Super-resolution and Image restoration }  
	
    
    Considering a low resolution and noisy image, several techniques have been proposed to reproduce the higher resolution and less noisy image.
	Image super-resolution (SR) is a class of techniques used for resolution enhancement in images. DL-based algorithms, especially with deeply and fully convolutional networks, have achieved high performance in the problem of recovering the HR images from the LR image inputs. Recently, Super-resolution convolutional neural network (SRCNN) \cite{SRCNN} is proposed to map between LR/HR images in an end-to-end manner.
	Other than SR techniques, image restoration (IR) algorithms can be applied to remove/reduce the noise effect on an image. 
	Various models have been presented for IR in the literature. For instance, in \cite{ِDenoising}, a feed-forward denoising convolutional neural network (DnCNN) scheme is presented which has utilized the residual learning and batch normalization to speed up the training process.

	 
	 

\section{ChannelNet} \label{sec:chnet}
\subsection{Channel Image} \label{sec:chanelimage}
\begin{figure}
	 	\centering
	 	\includegraphics[width=4.5cm,height=4.5cm,keepaspectratio]{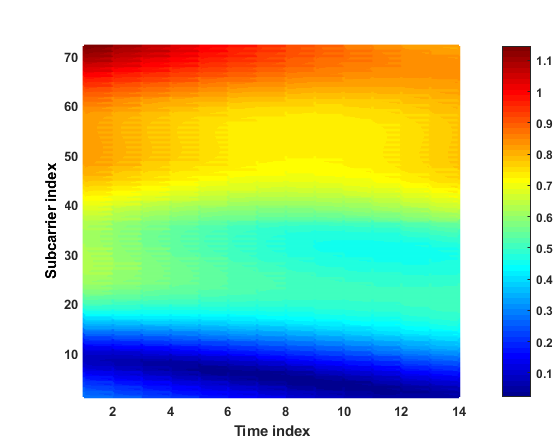}
	 	\caption{An example of normalized real/imaginary 2D-image for a sample channel time-frequency grid}
	 	\label{samplech}
	      \end{figure}

	 In this work, we focus on one link between a pair of Tx and Rx antennas, i.e., we have Single-input, Single-output (SISO) communication link. For this link, the channel time-frequency response matrix $\mathbf{H}$ (of size $N_S\times N_D$) between a transmitter and a receiver, which has complex values, can be represented as \textit{two 2D-images} (one 2D-image for real values and another one for imaginary values). An example of the normalized real/imaginary 2D-image for a sample channel time-frequency grid with $N_D = 14$ time slots and $N_S = 72$ subcarriers (based on Long-Term Evolution (LTE) standard) is shown in \figurename \ref{samplech}.
	 \subsection{Network Structure}

\begin{figure}
	 	\centering
	 	\includegraphics[width=6.5cm,height=7.8cm,keepaspectratio]{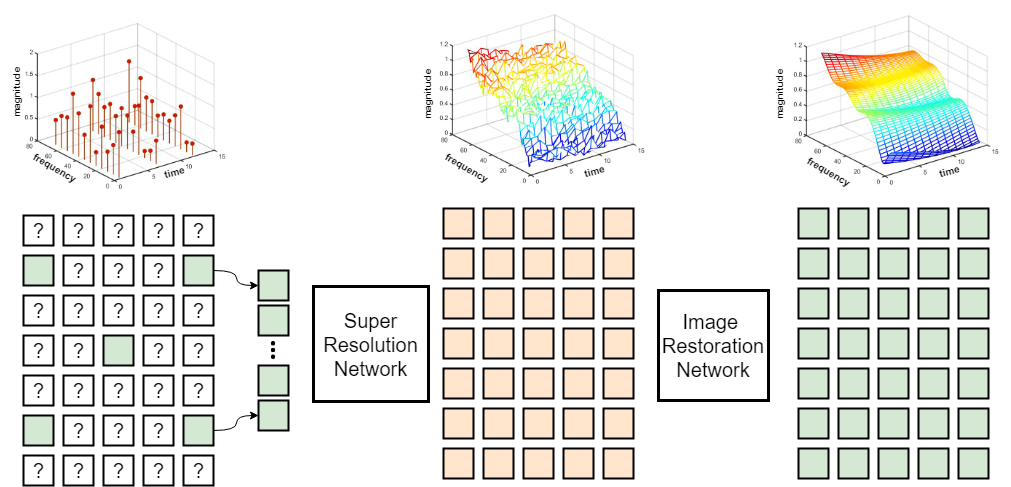}
	 	\caption{The proposed pipeline for DL-based channel estimation}
        \label{pipeline}
	\end{figure}

 The overview of the proposed pipeline for DL-based channel estimation, named ChannelNet, is illustrated in \figurename \ref{pipeline}. The goal is to estimate the whole time-frequency of the channel using the transmitted pilots. Similar to LTE standard, Lattice-type pilot arrangement has been used for pilot transmission. 
	 
The estimated value of the channel at the pilot locations $\hat{\textbf{h}}_{p}^{\mathrm{LS}}$ (which might be noisy) is considered as the LR and noisy version of the channel image. To obtain the complete channel image a two stage training approach is presented:
	 \begin{itemize}
	 	\item In the first stage, an SR network is implemented which takes $\hat{\textbf{h}}_{p}^{\mathrm{LS}}$ as the vectorized low resolution input image (once the real-part and then the imaginary-part) and estimates the unknown values of  channel response $\mathbf{H}$.
	 	\item In the second stage to remove the noise effects, a denoising IR network is cascaded with the SR network. 
	 \end{itemize}

	 For the \textit{SR} and \textit{IR}, we have used \textit{SRCNN} \cite{SRCNN} and \textit{DnCNN} \cite{ِDenoising}, respectively. Due to page limitation, we cannot show their structure pictorially. At a high level though, \textit{SRCNN} first uses an interpolation scheme to find the approximate values of the high resolution image (channel) and afterwards, improves the resolution using a three-layer convolutional network. The first convolutional layer uses 64 filters of size  $9 \times 9$ and the second layer uses 32 filters of size $1 \times 1$, both followed by ReLu activation. The final layer uses only one filter of size $5 \times 5$ to reconstruct the image. \textit{DnCNN} (details in \cite{ِDenoising}) is a residual-learning based network which composed of 20 convolutional layers. The first layer uses 64 filters of size $3 \times 3 \times 1$ followed by a ReLU. Each of the succeeding 18 convolutional layers uses 64 filters of size $3 \times 3 \times 64$ followed by batch-normalization and ReLU. The last layer uses one $3 \times 3 \times 64$ filter to reconstruct the output.

   	 \subsection{Training}

   	 Lets denote the set of all network parameters by $\Theta =\{\Theta_S,\Theta_R\}$, where the $\Theta_S$ and $\Theta_R$ denote the set of parameter values for SR and IR networks, respectively. The input to the ChannelNet is the pilot values vector $\hat{\textbf{h}}_{p}^{\mathrm{LS}}$ and the output is the estimated channel matrix is denoted as $\mathbf{\hat{H}}$:
   	 $$
   	 \mathbf{\hat{H}}=f(\Theta;\hat{\textbf{h}}_{p}^{\mathrm{LS}}) = f_R(f_S(\Theta_S;\hat{\textbf{h}}_{p}^{\mathrm{LS}});\Theta_R),$$
   	 where $f_S$ and $f_R$ are the SR and IR functions, respectively. 
   	 
	 The total loss function of the network is the Mean square error (MSE) between the estimated and the actual channel responses calculated as follows:
   	 
   	 	 \begin{equation} \label{eq:cost}
   	 	 C = \frac{1}{\|\cal{T}\|}  \sum_{\textbf{h}_{p} \in \cal{T}}\|f(\Theta;\hat{\textbf{h}}_{p}^{\mathrm{LS}})-\textbf{H}\|_2^2,
   	 	 \end{equation}
   	 	 where $\cal{T}$ is the set of all training data and $H$ is the perfect channel. In \eqref{eq:cost}, $\|\cal{T}\|$ is the size of the training set.
   	 
   	 To simplify the training process, we use a two stage training algorithm. Where in the first stage we 
   	 minimize  the the loss of the SR network, $C_1$ :
   	 
   	 \begin{equation}
   	 C_1 = \frac{1}{\|\cal{T}\|}  \sum_{\textbf{h}_{p} \in \cal{T}} \|\textbf{Z}-\textbf{H}\|_2^2,
   	 \end{equation}
   	 where $\textbf{Z} = f_S(\Theta_S;\hat{\textbf{h}}_{p}^{\mathrm{LS}})$ is the output of the SR network. 

   	 In the second stage, we freeze the weights of the SR network and find the parameters of the denoising network by defining $\hat{\textbf{H}} = f_R(\textbf{Z};\Theta_D)$ and  minimizing the loss function $C_2$ :
   	 
   	 \begin{equation}
   	 C_2 = \frac{1}{\|\cal{T}\|}  \sum_{\textbf{h}_{p} \in \cal{T}}\|\hat{\textbf{H}}-\textbf{H}\|_2^2,
   	 \end{equation}


   	 Note that, similar to image-based techniques, the optimal weights of the network is dependent on the value of the SNR; thus, to have a complete solution we have to re-train the network for each SNR value. This approach is practically impossible to implement because the SNR value is continuous. Fortunately, however, as the results in section \ref{sec:sim} demonstrates, training networks for a few SNR values (in our case only two values) can still lead to a good performance. 

	 \section{Simulation Results}	 \label{sec:sim}	 
	 
	 In this section we train the network and evaluate the MSE over a range of SNRs and compare the results with the widely used baseline algorithms. 

	 We consider a single antenna at the transmitter and at the receiver. For the channel modeling and pilot transmission, we have used widely used LTE simulator developed by university of Vienna,  Vienna LTE-A simulator \cite{Mehlführer2011}. Keras and Tensorflow using a GPU backend are used for implementation of our proposed scheme. For SR and IR networks, the training rate is set to 0.001 with batch size of 128 and at most 500 iterations. The training, testing and validation sets consist of 32000, 4000 and 4000 channels, respectively\footnote{ Source code : \url{https://github.com/Mehran-Soltani/ChannelNet}}. As in LTE, in our simulations, each frame consists of 14 time slots with 72 subcarriers. For the wireless channel models of Vehicular-A (VehA) and \textcolor{black}{SUI5 (a model with long delay spread)}  with carrier frequency of 2.1 GHz, bandwidth of 1.6 MHz and UE (user equipment) speed of 50 km/h,  are considered.
	 
	 
	 To see the performance, we have compared the accuracy of channel estimation for the proposed method with that of three state-of-the-art algorithms i.e. ideal MMSE, estimated MMSE and ideal ALMMSE \cite{ALMMSE_CO} when 48 pilots are used in each frame. The MSE between the estimated and the actual channel realization is considered as the performance metric. 
	 
\begin{figure}
	 	\centering
	 	\includegraphics[width=7cm,height=4.2cm,keepaspectratio]{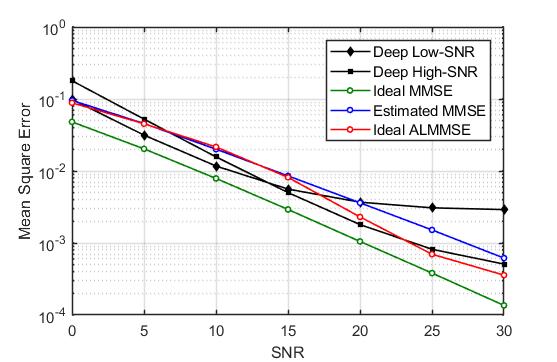}
	 	\caption{Channel Estimation MSE in terms of SNR for VehA channel model.}
        \label{result_48}
	 \end{figure}

	 The results for VehA is presented in  \figurename \ref{result_48}.  Note that the ideal MMSE has the best performance and gives a lower bound of the achievable MSE as the channel correlation matrix should be known fully (without any error) which is not a valid assumption in practical applications. Estimated MMSE tries to estimate the correlation matrix based on received signals and ideal ALMMSE is an approximate counterpart of ideal MMSE (but still has the complete knowledge of the channel statistics). 
	 
	  In \figurename \ref{result_48}, it is demonstrated that for low SNR values, the proposed ChannelNet trained at the SNR value of 12dB (denoted by deep low-SNR)  has comparable performance with the ideal MMSE and has a better performance than the ideal ALMMSE and the estimated MMSE.  Additionally, it can be observed that after around a mid SNR value, the performance of the network trained at the SNR value of 22dB (denoted by deep high-SNR) is going to be better than the deep low-SNR. 

	  So, we divide the SNR range into two regions. When the SNR value is low, we estimate the channel by deep low-SNR network, and beyond a threshold, deep high-SNR network is used. It can be observed that for SNR values higher than 23 dB, the performance of deep high-SNR is going to fail again and another network has to be trained; though as long as the SNR is below 20 dB, the two generated networks are sufficient.
	  
\begin{figure}
	 	\centering
	 	\includegraphics[width=7cm,height=4.2cm,keepaspectratio]{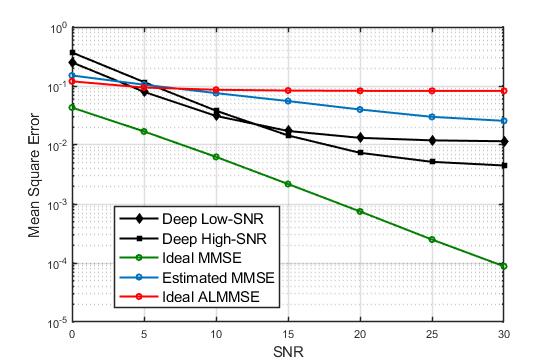}
	 	\caption{Channel estimation MSE in terms of SNR for SUI5 channel model.}
        \label{SUI}
	 \end{figure}

	  \textcolor{black}{MSE results associated with SUI5 model are depicted in \figurename \ref{SUI}. In general, due to higher complexity of channel, all schemes show lower performance compared to the VehA model. More interestingly, after SNR value of 5dB, we can observe that schemes like ALMMSE and Estimated MMSE degraded significantly while the proposed deep model can still discover the underlying statistics and gets to an acceptable MSEs. As we expect, Ideal-MMSE has the best performance but it is not achievable in practical scenarios as it needs full knowledge of the correct channel statistics.} 
	 
\begin{figure}
	 	\centering
	 	\includegraphics[width=7cm,height=4.2cm,keepaspectratio]{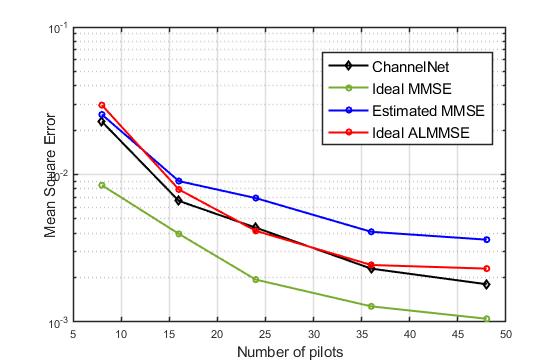}
	 	\caption{Mean square error for channel estimation in terms of pilot number.}
        \label{pilot length}
	 \end{figure}

	 To show the performance of the proposed algorithm, considering VehA channel model, results of simulations for different number of pilots at the SNR level of 20dB are depicted in \figurename \ref{pilot length}. As can be seen, the ChannelNet, trained at that specific value of SNR, outperforms the Estimated MMSE and Ideal ALMMSE methods and it is comparable to Ideal MMSE.

	 
 \section{Conclusion} \label{sec:conc}
In this paper, we presented ChannelNet, our initial DL-based algorithm for channel estimation in communication systems. In this method, we have considered the time-frequency response of a fading channel as a 2D-image and applied SR and IR algorithms to find the whole channel state based on the pilot values. The results show that the performance of ChannelNet is highly competitive with the MMSE algorithm. The two-step network training procedure has been presented and we also discussed how multiple ChannelNets should be used to best estimate the channel. 

\bibliography{ref}
\bibliographystyle{ieeetr}

\end{document}